\documentclass[onecolumn,12pt,superscriptaddress,nofootinbib]{revtex4}
\pdfoutput=1
\usepackage[latin1]{inputenc}
\usepackage[english]{babel}
\usepackage{amssymb}
\usepackage{amsmath}
\usepackage{amsthm}
\usepackage[]{graphicx}
\usepackage[]{subfigure}
\usepackage{tensor}
\usepackage{float}
\usepackage{color}
\usepackage{cancel}
\usepackage{setspace}
\usepackage{fancyhdr}
\usepackage{framed}
\usepackage{braket}
\usepackage{tikz}
\usepackage{comment}
\DeclareMathOperator{\sech}{sech}
\usepackage[bookmarks,linktocpage, colorlinks=true, plainpages = false, citecolor = red,  linkcolor=blue, urlcolor = magenta, filecolor = blue]{hyperref}

\usepackage{mathrsfs}
\usepackage[title]{appendix}
\numberwithin{equation}{section}

\begin{document}

\allowdisplaybreaks

\title{What can we learn from the conformal noninvariance of the Klein-Gordon equation?}
\vspace{.3in}

\author{F. Hammad}
\email{fhammad@ubishops.ca}
\affiliation{Department of Physics and Astronomy, Bishop's University, 2600 College Street, Sherbrooke, QC, J1M~1Z7
Canada}
\affiliation{Physics Department, Champlain 
College-Lennoxville, 2580 College Street, Sherbrooke,  
QC, J1M~2K3 Canada}
\affiliation{D\'epartement de Physique, Universit\'e de Montr\'eal,\\
2900 Boulevard \'Edouard-Montpetit,
Montr\'eal, QC, H3T 1J4
Canada} 

\author{P. Sadeghi} \email{psadeghi20@ubishops.ca} 
\affiliation{Department of Physics and Astronomy, Bishop's University, 2600 College Street, Sherbrooke, QC, J1M~1Z7
Canada}

\author{N. Fleury} \email{flen2202@usherbrooke.ca} 
\affiliation{Department of Physics, Universit\'e de Sherbrooke, Sherbrooke, QC, J1K~2X9
Canada}

\author{A. Leblanc} \email{leba3108@usherbrooke.ca} 
\affiliation{Department of Physics, Universit\'e de Sherbrooke, Sherbrooke, QC, J1K~2X9
Canada}

\begin{abstract}
It is well known that the Klein-Gordon equation in curved spacetime is conformally noninvariant, both with and without a mass term. We show that such a noninvariance provides nontrivial physical insights at different levels, first within the fully relativistic regime, then in the nonrelativistic regime leading to the Schr\"odinger equation, and then within the de Broglie-Bohm causal interpretation of quantum mechanics. The conformal noninvariance of the Klein-Gordon equation coupled to a vector potential is confronted with the conformal invariance of Maxwell's equations in the presence of a charged current. The conformal invariance of the non-minimally coupled Klein-Gordon equation to gravity is then examined in light of the conformal invariance of Maxwell's equations. Finally, the consequence of the noninvariance of the equation on the Aharonov-Bohm effect in curved spacetime is discussed.
\end{abstract}


\maketitle

\section{Introduction}\label{Intro}
It is very well known that the Klein-Gordon equation (KGE) describing the dynamics of relativistic spinless quantum particles in curved spacetime is conformally noninvariant, both with and without a mass term (see, e.g., the textbooks \cite{QFCS1,QFCS2} or the review paper \cite{ConformalIssue}). By conformal invariance in curved spacetime, one means the preservation of the form of the mathematical equations under a conformal mapping of the spacetime. By such a mapping, we mean a rescaling of the time and space intervals without changing the coordinates used to describe spacetime events. In this sense, such a transformation, usually called a Weyl conformal transformation, is a physical transformation of the spacetime and of its content in mass, energy and pressure in contrast to a coordinate conformal transformation in a flat spacetime (see, e.g., Refs.\,\cite{Particles+Fields,PLB}).

Investigating the consequences of a Weyl conformal transformation on spacetime usually leads to very interesting insights about physical phenomena related to the dynamics of spacetime, ranging from the concept of quasi-local masses in general relativity \cite{Prain,Misner-Sharp,HH} to the physics of wormholes \cite{Faraoni,BHWormhole,WormholeEC} and black holes \cite{Nielsen0,Nielsen2,Nielsen3,Nielsen4,BHThermo,FDV,BHWormhole,STThermo,Teleparallel}. 
Unfortunately, more often than not, the noninvariance of the KGE under a Weyl conformal transformation is mentioned in the literature as a pathology or, at best, simply as one of the arguments for introducing a non-minimal coupling of the Klein-Gordon scalar field to gravity \cite{QFCS1,QFCS2,QFCS3}. As we shall see in this paper, however, an interesting and rich physics is actually hiding behind such a noninvariance of the equation.

For research focused on solving the KGE in curved spacetime, see, e.g., the more recent works in Refs.\,\cite{Landry,KGCurved2018,GravityLandauI,GravityLandauII} and the references therein. For works focused instead on bringing to light the influence of gravity on the {\it quantum behavior} of a particle as the latter interacts with the curved spacetime, we refer the reader to a handful of references we are aware of that were devoted to a close investigation of such a problem \cite{Papini1965,Cai1989,Cai1990,QPhaseI,QPhaseIII,QPhaseII,Nandi2009,DiracKGCurved2011}. In the latter references, the main focus was the gravitational effect on the phase $\phi$ of the wavefunction of a particle as the latter propagates in a curved spacetime. One of our goals in this paper is to extend such an investigation using the Weyl conformal transformation tool. 

It was suggested very early on in Ref.\,\cite{Stodolski} that gravity must manifest itself through the metric $g_{\mu\nu}$ of spacetime by inducing a phase $\phi=\frac{1}{\hbar}\int g_{\mu\nu}\,p^\mu{\rm d}x^\nu$ on any particle of 4-momentum $p^\mu$, where $\hbar$ is Planck's constant and the integral is to be performed over the particle's path. This expression of the phase is, in fact, the basis behind the modern use of neutrino interferometry (see, e.g., Ref.\,\cite{NeutrinoInCS} and the very recent work \cite{NWPG} and the references therein) and other quantum objects \cite{Josephson,COWBall,Science} and quantum phenomena \cite{QHE} for probing gravity. According to this expression, it seems that in order to find the effect of a Weyl transformation on the phase $\phi$ one only needs to find the effect of such a transformation on the 4-momentum $p^\mu$ by relying on the relativistic and conformally invariant formula $g_{\mu\nu}p^\mu p^\nu=-m^2c^2$. As we shall see, this is not the case.

It is partly in this regard that the hydrodynamic de Broglie-Bohm approach to quantum mechanics \cite{Broglie,Bohm,Holland,Solvey27} becomes relevant to our present investigation. Recall, indeed, that in such an approach, the real spacetime-dependent phase $\frac{1}{\hbar}S(x)$ of the particle's wavefunction $\varphi(x)$ appears when the latter is expressed using the ansatz $\varphi(x)=R(x)e^{\frac{i}{\hbar}S(x)}$, with a real amplitude $R(x)$. However, in such an approach the real function $S(x)$ is related to the amplitude $R(x)$ via an analog of the classical Hamilton-Jacobi differential equation. Therefore, one can no longer separate the effects of a Weyl transformation on the phase of the wavefunction from the effects on the amplitude of the latter.

The remainder of this paper is organized as follows. In Sec.\,\ref{Relativistic}, we examine the implications of the conformal noninvariance of the KGE using the fully relativistic regime of the equation. We first conduct our analysis using the concept of current, then we use the Lagrangian formalism to consolidate our results. The right Weyl transformation of $\varphi(x)$ that will be used in the rest of this paper is derived in that section and is shown to be unique. The physical interpretation of our results are discussed. In Sec.\,\ref{NonRelativistic}, we examine the physical consequences of such a noninvariance within the nonrelativistic limit of the equation. The significance of our results in relation to the foundations of quantum mechanics are discussed. In Sec.\,\ref{dBBApproach}, we look at the issue from the de Broglie-Bohm hydrodynamics approach to quantum mechanics. In Sec.\,\ref{LinkWithMaxwell}, we revisit the conformal invariance of Maxwell's equations in light of this noninvariance of the KGE. In Sec.\,\ref{Sec:VectorConstruction}, we extract the conformally invariant non-minimally coupled KGE from a conformally invariant vector field Lagrangian. In Sec.\,\ref{AharonovBohm}, we discuss the implications of the Weyl transformation on the Aharonov-Bohm effect and its link with the conformal noninvariance of the KGE. We conclude this paper with a brief summary and discussion section.
\section{In the relativistic regime}\label{Relativistic}
\subsection{Using the conserved current}\label{UsingCurrent}
In the flat Minkowski spacetime, the KGE for spin-0 particles of mass $m$ reads, $\left(\eta^{\mu\nu}\partial_\mu\partial_\nu-m^2c^2/\hbar^2\right)\varphi=0$. Here, $\eta_{\mu\nu}$ stands for the Minkowski metric, and the spacetime signature we adopt in this paper is $(-,+,+,+)$.
When going to a curved spacetime of metric $g_{\mu\nu}$, the KGE is built using the minimal substitution prescription \cite{QFCS1,QFCS2} of turning partial derivatives $\partial_\mu$ into covariant derivatives $\nabla_\mu$ and replacing the flat metric $\eta_{\mu\nu}$ by the general metric $g_{\mu\nu}$. In curved spacetime, the KGE reads
\begin{equation}\label{KGCST}
\left(g^{\mu\nu}\nabla_\mu\nabla_\nu-\frac{m^2c^2}{\hbar^
2}\right)\varphi=0.
\end{equation}

The problem one encounters when attempting to interpret the KGE as describing a single quantum particle is that of a possible negative energy and that of a non-positive definite probability density. Thanks to the Feynman propagator prescription \cite{BD}, however, one overcomes such limitations by interpreting particles with negative energy as anti-particles propagating backward in time \cite{Stuckelberg,Feynman}. Therefore, as long as the creation and annihilation of particles can be ignored in the physical process one is interested in, the machinery of quantum field theory that consists in viewing $\varphi(x)$ as a scalar field expandable in terms of creation and annihilation operators is not required. This is why the $\pi$ and $K$ mesons, for example, are successfully described at low energy with the propagator approach based on the KGE \cite{BD}. In a scattering process, these scalar particles are indeed stable particles within their lifetime and are thus represented as free initial and final wavefunctions. It is therefore in this approximation that we can also describe the behavior of a spinless particle obeying the KGE under the Weyl transformation. As such, the ambiguity in defining positive and negative frequencies to be attached to the creation and annihilation operators need not concern us here.

In fact, the other way of thinking of the negative probability densities that arise from the KGE is provided by considering a complex wavefunction $\varphi(x)$. By multiplying by the elementary electric charge $e$ the conserved 4-current that emerges from Eq.\,(\ref{KGCST}), one obtains the following covariantly conserved charged current density \cite{Greiner}
\begin{equation}\label{KGCurrent}
    j_\mu(x)=\frac{i\hbar e}{2m}\left(\varphi^*\nabla_\mu\varphi-\varphi\nabla_\mu\varphi^*\right).
\end{equation}
A negative probability density current can then simply be interpreted as representing a negatively charged electric current density, regardless of whether the wavefunction describes a single-particle or a particle-antiparticle state.

Let us now deform spacetime {\it \`a la} Weyl by multiplying the spacetime metric $g_{\mu\nu}(x)$ at each point of the spacetime by the positive, spacetime-dependent, and everywhere smooth conformal factor $\Omega^2(x)$. A new spacetime emerges then with a metric $\tilde{g}_{\mu\nu}(x)$ that is related to the old metric $g_{\mu\nu}(x)$ through the equation, $\tilde{g}_{\mu\nu}(x)=\Omega^{2}g_{\mu\nu}(x)$ \cite{Wald}. Our first goal in this section is to derive the KGE that the wavefunction $\varphi(x)$ of the old spacetime would obey in the new spacetime. Using the definition of the Christoffel symbols in terms of the metric and its derivatives \cite{Wald}, we easily check that for any conformal factor $\Omega$, we have that $g^{\mu\nu}\Gamma_{\mu\nu}^\lambda=\Omega
^{2}\tilde{g}^{\mu\nu}\tilde{\Gamma}_{\mu\nu}
^\lambda+2\Omega\tilde{g}^{\rho\lambda}\partial
_\rho\Omega$ \cite{Wald,ConformalIssue}. Using this identity, we arrive at the following equation obeyed by the old wavefunction $\varphi(x)$ in the new spacetime:
\begin{equation}\label{ConfKG}
\left(\tilde{g}^{\mu\nu}\tilde{\nabla}_\mu\tilde{\nabla}_\nu-\frac{\tilde{m}^2c^2}{\hbar^2}\right)\varphi=2\frac{\tilde{\nabla}^\mu\Omega}{\Omega}\tilde{\nabla}_\mu\varphi.
\end{equation}
In obtaining this equation, we have replaced the old constant mass $m$ as perceived in the old spacetime by the new mass $\tilde{m}$ as perceived in the new spacetime. The two masses are related by $\tilde{m}=m/\Omega$. This conformal transformation of mass can be arrived at both classically and quantum mechanically.

Classically, the Weyl transformation of mass arises, for example, from the simple way the energy-momentum tensor $T^{\mu\nu}$ transforms when it is traceless. In fact, the energy-momentum tensor extracted from an invariant action transforms as $\tilde{T}_\mu\,^\nu=\Omega^{-4}T_\mu\,^\nu$ \cite{ConformalIssue}. On the other hand, for an electromagnetic field, the energy-momentum tensor is traceless and its component $T_0\,^0$ is simply an energy-density \cite{LandauLifshitz}. Therefore, energy density $\rho$ should transform as $\tilde{\rho}=\Omega^{-4}\rho$. Given that the physical volume $V$ of space transforms as $\tilde{V}=\Omega^3 V$, we deduce that mass, like energy, has to transform as $\tilde{m}=m/\Omega$. For another classical argument, see Ref.\,\cite{Fulton}.   

Quantum mechanically, the simplest argument for the above transformation of mass can be formulated as follows. The Compton wavelength of a particle of mass $m$ is defined by $\lambda_C=h/mc$. Given that $\lambda_C$ is a physical length in spacetime and given that it has nothing to do with the coordinates of such a spacetime, we conclude that it should transform as $\tilde{\lambda}_C=\Omega\lambda_C$, from which the transformation of mass follows.   

The result (\ref{ConfKG}) clearly shows the conformal noninvariance of the KGE. Furthermore, it clearly shows that it does not help to set the mass $m$ to zero in the equation to achieve a conformal invariance of the latter. From the result (\ref{ConfKG}), our first interesting physical interpretation of the effect of the Weyl transformation on the wavefunction already emerges. The old wavefunction $\varphi(x)$ is perceived in the new spacetime as a wavefunction obeying a KGE with a source term coupled to the gradient of the wavefunction. The source is, what we could call, the spacetime ``deformation current": $j_\mu^\Omega(x)=2\tilde{\nabla}_\mu\Omega/\Omega$. In fact, we see from this equation that the only way for the old wavefunction $\varphi(x)$ to also obey the KGE in the new spacetime, is to have the condition $\tilde{\nabla}^\mu\Omega\,\tilde{\nabla}_\mu\varphi=0$ satisfied. Physically, this means that only those waves propagating along the ``troughs'' of the deformed spacetime can also be solutions to the wave equation in the deformed spacetime. Such wavefunctions would in fact be insensitive to the deformations of spacetime (the particles avoid the geometrically induced force), for they make their way though the deformations by propagating along the least ``resistant'' paths.

Now, the next step is to assume that $\varphi(x)$ is also transformed by the Weyl rescaling in such a way that there is a linear relation between the new $\tilde{\varphi}(x)$ and the old $\varphi(x)$ written in the form $\varphi(x)=\Omega^{s}\tilde{\varphi}(x)$, for some real exponent $s$, called a conformal weight\footnote{Note that our conformal weight is here the negative of the one usually displayed in textbooks. The form we displayed here suits better our present analysis.}. We then find that $\tilde{\varphi}(x)$ obeys the following KGE in the new spacetime:
\begin{equation}\label{ConfKG+s}
\left(\tilde{\Box}-\frac{\tilde{m}^2c^2}{\hbar^2}\right)\tilde{\varphi}+\left[(s^2+3s)\frac{\tilde{\nabla}_\mu\Omega}{\Omega}\frac{\tilde{\nabla}^\mu\Omega}{\Omega}-s\frac{\tilde{\Box}\Omega}{\Omega}\right]\tilde{\varphi}=2(s+1)\frac{\tilde{\nabla}^{\mu}\Omega}{\Omega}\tilde{\nabla}_\mu\tilde{\varphi}.
\end{equation}
This equation clearly shows that there is no conformal weight $s$ for which $\tilde{\varphi}(x)$ would obey the usual KGE in the new spacetime for an arbitrary conformal factor $\Omega(x)$. 

Let us then assume a more general transformation and set $\varphi(x)=f(\tilde{\varphi},\tilde{\varphi}^*,\Omega)$, for an arbitrary functional $f$ which is at least of class $C^2$ in its arguments. Substituting this functional into the KGE (\ref{ConfKG}), the latter takes the form,
\begin{multline}\label{ConfKG+f}
f_{,\tilde{\varphi}}\,\tilde{\Box}\tilde{\varphi}-\frac{\tilde{m}^2c^2}{\hbar^2}f+f_{,\tilde{\varphi}\tilde{\varphi}}\tilde{\nabla}_\mu\tilde{\varphi}\tilde{\nabla}^\mu\tilde{\varphi}+2\left(f_{,\tilde{\varphi}\Omega}-\frac{f_{,\tilde{\varphi}}}{\Omega}\right)\tilde{\nabla}^{\mu}\Omega\tilde{\nabla}_\mu\tilde{\varphi}\\
+f_{,\tilde{\varphi}^*}\,\tilde{\Box}\tilde{\varphi}^*+f_{,\tilde{\varphi}^*\tilde{\varphi}^*}\tilde{\nabla}_\mu\tilde{\varphi}^*\tilde{\nabla}^\mu\tilde{\varphi}^*+2\left(f_{,\tilde{\varphi}^*\Omega}-\frac{f_{,\tilde{\varphi}^*}}{\Omega}\right)\tilde{\nabla}^{\mu}\Omega\tilde{\nabla}_\mu\tilde{\varphi}^*+2f_{,\tilde{\varphi}\tilde{\varphi}^*}\tilde{\nabla}_\mu\tilde{\varphi}\tilde{\nabla}^\mu\tilde{\varphi}^*\\
+\tilde{\Box}\Omega f_{,\Omega}+\tilde{\nabla}_\mu\Omega\tilde{\nabla}^\mu\Omega f_{,\Omega\Omega}-2\frac{\tilde{\nabla}_{\mu}\Omega\tilde{\nabla}^\mu\Omega}{\Omega}f_{,\Omega}=0.
\end{multline}
We see from this equation that no functional $f$ would allow $\tilde{\varphi}(x)$ to satisfy the KGE in the new spacetime without introducing extra constraints on $\tilde{\varphi}(x)$ besides the dynamics that would already be imposed on it by the KGE. Note that we did not consider here the possibility of a functional of the form $f(x,\tilde{\varphi},\tilde{\varphi}^*,\Omega)$, for it would entail a different analytic form at different spacetime points. From these results, it becomes then evident that one has to abandon not only the assumption of linearity of the transformation of $\varphi(x)$, but even the assumption of locality.

It turns out that the transformed KGE (\ref{ConfKG}) itself leads in a remarkably easy way to the sought-after Weyl transformation. In fact, multiplying both sides of Eq.\,(\ref{ConfKG}) by $\varphi^*$, then taking the complex conjugate of the resulting equation, and then subtracting such an equation from the equation before complex conjugation, we arrive at the following result:

\begin{equation}\label{ConfKGCurrent}
\tilde{\nabla}^\mu\left(\varphi^*\tilde{\nabla}_{\mu}\varphi-\varphi\tilde{\nabla}_{\mu}\varphi^*\right)=2\frac{\tilde{\nabla}^\mu\Omega}{\Omega}\left(\varphi^*\tilde{\nabla}_{\mu}\varphi-\varphi\tilde{\nabla}_{\mu}\varphi^*\right).
\end{equation}
This equation shows that in the new spacetime the ``Klein-Gordon'' current $J_\mu(\varphi)$, made of the old wavefunction $\varphi(x)$, and consisting of the content of the parentheses on both sides of the equation, is not covariantly conserved anymore. Eq.\,(\ref{ConfKGCurrent}) states that in the conformal spacetime we have $\tilde{\nabla}^\mu J_\mu(\varphi)=2\tilde{\nabla}^\mu\Omega\, J_\mu(\varphi)/\Omega$. In other words, $\varphi(x)$ is not freely propagating in the new spacetime, as we already saw it from a different perspective based on Eq.\,(\ref{ConfKG}). The current made of $\varphi(x)$ appears in the conformal frame as a current made of charged particles interacting with the deformation current $j_\mu^\Omega(x)=2\tilde{\nabla}_\mu\Omega/\Omega$.

It is now possible to see from Eq.\,(\ref{ConfKGCurrent}) why no local analytic functional of the form $\varphi=f(\tilde{\varphi},\tilde{\varphi}^*,\Omega)$ could be found in our previous attempts to relate $\varphi(x)$ to $\tilde{\varphi}(x)$. First, rewrite Eq.\,(\ref{ConfKGCurrent}) in the form $\tilde{\nabla}^\mu\left[\Omega
^{-2}J_{\mu}(\varphi)\right]=0$. This implies that the current $\Omega^{-2}J_\mu(\varphi)$ is covariantly conserved in the new spacetime. On the other hand, we know that in the new spacetime the current $\tilde{J}_\mu(\tilde{\varphi})$, made of a wavefunction $\tilde{\varphi}(x)$ satisfying the KGE, obeys the covariant conservation equation as well: $\tilde{\nabla}^\mu\tilde{J}_\mu(\tilde{\varphi})=0$. We therefore conclude that $\tilde{J}_\mu(\tilde{\varphi})$ and $\Omega^{-2}J_\mu(\varphi)$ must be proportional to each other for an arbitrary proportionality constant that we should take to be one to preserve unitarity under the Weyl transformation.

In fact, suppose that there exists in the new spacetime a 4-vector $\tilde{v}_\mu$ made of $\Omega$, $\tilde{\varphi}$, $\tilde{\varphi}^*$ and their derivatives, such that $\Omega^{-2}J_\mu(\varphi)=\tilde{J}_\mu(\tilde{\varphi})+\tilde{v}_\mu$, or, equivalently, that $J_\mu(\varphi)=\Omega^2\tilde{J}_\mu(\tilde{\varphi})+\Omega^2\tilde{v}_\mu$. Since we go from the conformal spacetime back to the original spacetime by using $\Omega^{-2}$, we deduce by iterating this identity that there must exist in the original spacetime a divergence-free 4-vector $v_\mu$ such that, $J_\mu(\varphi)=\Omega^2\left[\Omega^{-2}J_\mu(\varphi)+\Omega^{-2}v_\mu\right]+\Omega^2\tilde{v}_\mu$. This implies that we must have $v_\mu=-\Omega^2\tilde{v}_\mu$ for any $\Omega$. This, in turn, means that $v_\mu$ and $\tilde{v}_\mu$ cannot simply be proportional, respectively, to the currents $J_\mu(\varphi)$ and $\tilde{J}_\mu(\tilde{\varphi})$. Thereby, $\varphi$ and $\tilde{\varphi}$ would each obey one extra equation besides the KGE in their respective spacetimes. This contradicts our assumption. We conclude then that the only relation between $\varphi$ and $\tilde{\varphi}$ is given by $\tilde{J}_\mu(\tilde{\varphi})=\Omega^{-2}J_\mu(\varphi)$.   

Thus, the old wavefunction $\varphi$ is related to the new wavefunction $\tilde{\varphi}$ not through a local analytic functional $f$, but in a nonlocal way such that the following differential equation is satisfied:
\begin{equation}\label{PhiNonlocality}
\tilde{\varphi}^*\nabla_\mu\tilde{\varphi}-\tilde{\varphi}\nabla_\mu\tilde{\varphi}^*=\Omega^{-2}\left(\varphi^*\nabla_\mu\varphi-\varphi\nabla_\mu\varphi^*\right).
\end{equation}
Note that, as the covariant derivatives act on scalars, we preferred to display here the same operator $\nabla_\mu$ on both sides of the equation for a later convenience. 

Very important to note also is that at the heart of our result lies only the nonlocality of the wavefunction, not the linearity or the nonlinearity of the wave equation. In fact, if we perform the same operations that led us to the result (\ref{ConfKGCurrent}), but using a nonlinear version of the KGE, such as the generalized Gross-Pitaevskii equation in curved spacetime (see, e.g., Refs.\,\cite{Anandan,Suarez}) 
\begin{equation}
\Box\varphi+\frac{m^2c^2}{\hbar^2}\varphi+\frac{8\pi a_sm}{\hbar^2}|\varphi|^2\varphi=0,
\end{equation}
where $a_s$ is a constant, we would also arrive at the same expression as in Eq.\,(\ref{ConfKGCurrent}). 

\subsection{Using the Lagrangian formalism}\label{UsingLagragian}
Both sides of Eq.\,(\ref{PhiNonlocality}) involve a purely imaginary quantity. It might then seem odd that the link between the wavefunctions in the two spacetimes does not involve a fully complex equation. This might suggest that such a link could somehow be incomplete as it leaves out half of the information complex equations convey. Therefore, in order to look for such a possibility, we shall turn now to the Lagrangian formulation of the KGE and the associated energy-momentum tensor. The Lagrangian from which the KGE (\ref{KGCST}) emerges by variation with respect to $\varphi^*$, reads
\begin{equation}\label{PhiLagrangian}
\mathcal{L}=-\frac{\hbar^2}{2m}\left(\nabla_\mu\varphi^*\nabla^\mu\varphi+\frac{m^2c^2}{\hbar^2}\varphi^*\varphi\right).
\end{equation}
The energy-momentum tensor, defined by $T_{\mu\nu}=\frac{2}{\sqrt{-g}}\frac{\delta}{\delta g^{\mu\nu}}(\sqrt{-g}\mathcal{L})$, that we extract from this Lagrangian is expressed as follows:
\begin{equation}\label{TTensor}
T_{\mu\nu}=-\frac{\hbar^2}{2m}\left[\nabla_\mu\varphi^*\nabla_\nu\varphi+\nabla_\mu\varphi\nabla_\nu\varphi^*-g_{\mu\nu}\left(\nabla_\rho\varphi^*\nabla^\rho\varphi+\frac{m^2c^2}{\hbar^2}\varphi^*\varphi\right)\right].
\end{equation}
By using the KGE (\ref{KGCST}), we easily check that this energy-momentum tensor is covariantly conserved; that is $\nabla^\mu T_{\mu\nu}=0$. In order to extract the KGE for $\tilde{\varphi}$ with a mass $\tilde{m}$ in the conformal spacetime, the Lagrangian to use should read as follows:
\begin{equation}\label{TildePhiLagrangian}
\tilde{\mathcal{L}}=-\frac{\hbar^2}{2\tilde{m}\Omega}\left(\tilde{\nabla}_\mu\tilde{\varphi}^*\tilde{\nabla}^\mu\tilde{\varphi}+\frac{\tilde{m}^2c^2}{\hbar^2}\tilde{\varphi}^*\tilde{\varphi}\right).
\end{equation}
In Appendix \ref{Appendix}, we show that this Lagrangian gives the correct KGE and yields Eq.\,(\ref{ConfKG}).
The energy-momentum tensor that we extract from this Lagrangian reads,
\begin{equation}\label{TildeTTensor}
\tilde{T}_{\mu\nu}=-\frac{\hbar^2}{2\tilde{m}\Omega}\left[\tilde{\nabla}_\mu\tilde{\varphi}^*\tilde{\nabla}_\nu\tilde{\varphi}+\tilde{\nabla}_\mu\tilde{\varphi}\tilde{\nabla}_\nu\tilde{\varphi}^*-\tilde{g}_{\mu\nu}\left(\tilde{\nabla}_\rho\tilde{\varphi}^*\tilde{\nabla}^\rho\tilde{\varphi}+\frac{\tilde{m}^2c^2}{\hbar^2}\tilde{\varphi}^*\tilde{\varphi}\right)\right].
\end{equation}
It is also straightforward to check that this energy-momentum tensor is not covariantly conserved. In fact, by using the KGE we derive from the Lagrangian (\ref{TildePhiLagrangian}), we find that $\tilde{\nabla}^\mu\tilde{T}_{\mu\nu}=\tilde{m}c^2\tilde{\varphi}^*\tilde{\varphi}\tilde{\nabla}_\nu\Omega^{-1}\neq0$. This non-conservation of $\tilde{T}_{\mu\nu}$ (often encountered in conformally transformed energy-momentum tensors \cite{ConformalIssue}) can be understood here as being due to the deformed geometry and a geometrically induced force as discussed below Eq.\,(\ref{ConfKG}), or equivalently, as being due to the deformation current as discussed below Eq.\,(\ref{ConfKGCurrent}). Therefore, in contrast to the conservation of current, there is no additional link between the two wavefunctions that could be extracted from the conservation of the energy-momentum tensor in the original spacetime.

A very important remark is here in order. One might argue that we could still try the well-known relation $\tilde{T}^{\mu\nu}=\Omega^{-6}T^{\mu\nu}$ \cite{ConformalIssue}. Unfortunately, this relation between the two energy-momentum tensors is valid only for the case of a conformally invariant action. Indeed, the derivation of such a relation requires $\sqrt{-g}\mathcal{L}=\sqrt{-\tilde{g}}\tilde{\mathcal{L}}$ \cite{ConformalIssue}. It can, however, easily be seen by referring to Eqs.\,(\ref{PhiLagrangian}) and (\ref{TildePhiLagrangian}) that imposing such an identity between the two Lagrangians leads to an inconsistent result with the one that would be obtained from $\tilde{T}^{\mu\nu}=\Omega^{-6}T^{\mu\nu}$ based on Eqs.\,(\ref{TTensor}) and (\ref{TildeTTensor}). Therefore,  the Lagrangians (\ref{PhiLagrangian}) and (\ref{TildePhiLagrangian}) cannot yield an invariant action for $\varphi(x)$ and cannot be used to build a link between the two wavefunctions. 

The relativistic treatment in this section allowed us to interpret the noninvariance of the KGE in terms of nonlocal transformations of $\varphi(x)$ and in terms of geometric-deformation currents that are all consistent with the non-conservation of the energy-momentum tensor in the new spacetime. In the next section, we examine the issue within nonrelativistic physics.
\section{In the nonrelativistic regime}\label{NonRelativistic}
\subsection{A worked out example}\label{Subsec:WorkedOut}
In the nonrelativistic regime, {\it i.e.}, when the energy of the particle is $E=mc^2+\mathcal{E}$ such that $mc^2\gg\mathcal{E}$, it is well known that the KGE in Minkowski spacetime reduces to the Schr\"odinger equation for a free particle \cite{Greiner}. We would like now to find the form of the Schr\"odinger equation in a conformal spacetime. 

According to Eq.\,(\ref{ConfKG}), the wavefunction $\tilde{\varphi}(x)$ that satisfies the KGE in a spacetime of metric $g_{\mu\nu}=\Omega^2\eta_{\mu\nu}$, obtained by conformally transforming the Minkowski spacetime, obeys in the latter the following equation:
\begin{equation}\label{ConfMink}
\left(\eta^{\mu\nu}\partial_\mu\partial_\nu-\frac{m^2c^2}{\hbar^2}\right)\tilde{\varphi}=-2\frac{\partial^\mu\Omega}{\Omega}\partial_\mu\tilde{\varphi}.
\end{equation}
In fact, the original-spacetime operator $\tilde{g}^{\mu\nu}\tilde{\nabla}_\mu\tilde{\nabla}_\nu-\frac{\tilde{m}^2c^2}{\hbar^2}$ acting on $\tilde{\varphi}$ can be written as $\Omega^{-2}\eta^{\mu\nu}\partial_\mu\partial_\nu-2\Omega^{-1}\eta^{\mu\nu}\partial_\mu\Omega^{-1}\partial_\nu-\Omega^{-2}\frac{m^2c^2}{\hbar^2}$, which leads to Eq.\,(\ref{ConfMink}). We will follow now the usual steps \cite{Greiner} for extracting the Schr\"odinger equation from this equation. It is, indeed, always possible to choose, in analogy with a linear gravitational field \cite{COWReview}, a deformation factor $\Omega(x)$ such that the ratio $\partial_\mu\Omega/\Omega$ can be treated as a small perturbation in the region of interest. 

To examine the nonrelativistic behavior to which $\tilde{\varphi}$ leads in the Minkowski spacetime, let us then set $\tilde{\varphi}(x)=\tilde{\psi}(x)e^{-\frac{i}{\hbar} mc^2t}$. Note that this ansatz involves a selection of a preferred time tailored to fit that of Minkowski using the constant mass $m$ as perceived in the latter, {\it i.e.}, we made a choice of a particular spatial slicing of the spacetime based on the unique possibility offered to us by the flat Minkowski spacetime. As we will see in Sec.\,\ref{Subsec:Schrodinger}, however, there is a heavy price to pay because of this. Inserting such an ansatz into Eq.\,(\ref{ConfMink}), the latter becomes
\begin{equation}\label{ConfSchro}
\left(i\hbar-\frac{\hbar^2}{mc^2}\frac{\dot\Omega}{\Omega}\right)\partial_t\tilde{\psi}=-\frac{\hbar^2}{2m}\nabla^2\tilde{\psi}-\frac{\hbar^2}{m\Omega}{\boldsymbol{\nabla}}\Omega\cdot{\boldsymbol{\nabla}}\tilde{\psi}-\frac{i\hbar\dot\Omega}{\Omega}\tilde{\psi}.
\end{equation}
This equation clearly displays a non-Hermitian Hamiltonian acting on the wavefunction $\tilde{\psi}$, a result which already alludes to a physical non-equivalence between the two spacetimes. In addition, we notice that, had we used the conformal mass instead and set $\tilde{\varphi}(x)=\tilde{\psi}(x)e^{-\frac{i}{\hbar}{\tilde m}c^2t}$, this issue would not have been solved. 

For simplicity, and in order to gain a more intuitive picture of the actual physics when the resulting Hamiltonian is Hermitian, we will choose in the remainder of this section a time-independent conformal factor $\Omega(x)$. Let us choose the conformal factor to depend only on the $z$-direction and to be given by $\Omega(z)=e^{\tanh\frac{z}{z_0}}$, for some arbitrary scale $z_0$. 

\begin{figure}[H]
    \centering
    \includegraphics[scale=0.53]{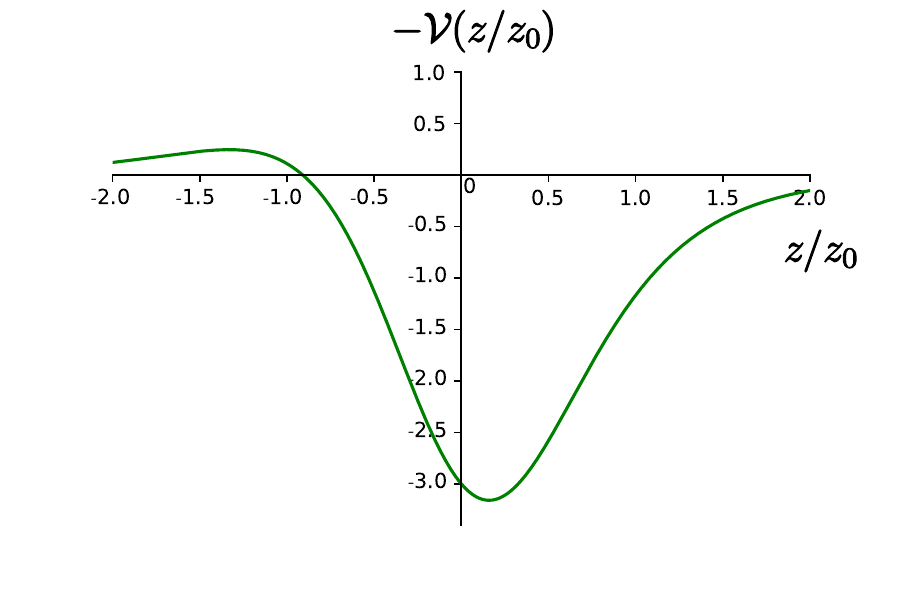}
    \caption{The shape of the function $-\mathcal{V}(z/z_0)$, where $\mathcal{V}(z/z_0)$ represents the content of the parentheses in Eq.\,(\ref{PsizEquation}), shows a potential well with a finite and smooth potential barrier on one side along the $z$-direction.}
    \label{fig:Potential}
\end{figure}

With such a conformal factor, Eq.\,(\ref{ConfSchro}) reduces to a time-independent equation with an extra term representing the interaction of the particle with the deformation of spacetime. Plugging the explicit expression of $\Omega(z)$ into Eq.\,(\ref{ConfSchro}), the latter takes the following form:
\begin{equation}\label{OmegazSchro}
-\frac{\hbar^2}{2m}\left(\partial^2_x+\partial^2_y+\partial^2_z\right)\tilde{\psi}-\frac{\hbar^2}{mz_0\cosh^2(z/z_0)}\,\partial_z\tilde{\psi}=\mathcal{E}\tilde{\psi}.
\end{equation}
This equation admits a solution which is separable in the coordinates of space. Therefore, seeking a solution of the form $\tilde{\psi}(x,y,z)=\tilde{\psi}_1(x)\tilde{\psi}_2(y)\tilde{\psi}_3(z)$, the equation splits into three independent equations such that the energy $\mathcal{E}$ on the right-hand side of the equation can be decomposed into $\mathcal{E}=\mathcal{E}_1+\mathcal{E}_2+\mathcal{E}_3$. Along the $x$- and $y$-direction, the equation describes a free particle. Along the $z$-direction, however, the equation for $\tilde{\psi}_3(z)$ is a more involved second-order differential equation which can be reduced to a Schr\"odinger equation with a simple redefinition of the wavefunction. We perform the redefinition $\tilde{\psi}_3(z)=e^{-\tanh\frac{z}{z_0}}\chi(z)$, so that we obtain an equation in $\chi(z)$ that reads
\begin{equation}\label{PsizEquation}
-\frac{\hbar^2}{2m}\partial^2_z\chi-\frac{\hbar^2}{2mz_0^2}\left(3\sech^4\tfrac{z}{z_0}+2\tanh \tfrac{z}{z_0}\sech^2\tfrac{z}{z_0}\right)\,\chi=\mathcal{E}_3\chi.
\end{equation}
This equation is indeed just a Schr\"odinger equation for the wavefunction $\chi(z)$ with a specific interaction potential. Note that a normalizable wavefunction $\chi(z)$ is sufficient to guarantee a normalizable wavefunction $\tilde{\psi}_3(z)$, for the multiplicative factor $e^{-\tanh\frac{z}{z_0}}$ is bounded for all values of $z$. In order to extract the physics hiding behind Eq.\,(\ref{PsizEquation}), we plotted in Figure\,\ref{fig:Potential} the shape of the potential as given by the variation in $z$ of the function $-\mathcal{V}(z/z_0)$, where $\mathcal{V}(z/z_0)$ represents the content inside the parentheses in Eq.\,(\ref{PsizEquation}).

This plot shows that the particle encounters a potential well with a finite and smooth potential barrier on one side along the $z$-direction. For positive values of $z$, the potential has the shape of a deep ravine displaying a negative minimum close to the origin, whereas for negative values of $z$ the potential has the shape of a smooth hill of positive values, which gradually becomes flat away from the origin. The consequence of this is that if the particle happens to propagate along the $z$-direction above the origin $z=0$, it could be trapped at the bottom of the well where it undergoes simple harmonic motion. The energy of the particle becomes then quantized and not continuous anymore. On the other hand, due to the presence of the positive potential hill below the origin $z=0$, the particle can only propagate downwards along the $z$-direction if it has an energy greater than the maximum potential of the hill. Therefore, for the particle to propagate freely along the $z$-direction its energy has to be larger than $-\hbar^2\mathcal{V}(\sigma)/2mz_0^2$, where $\sigma$ is the negative solution to the equation $12\tanh\sigma+4\sinh^2\sigma-2=0$, obtained from the first derivative of $\mathcal{V}(\mathcal{\sigma})$. Otherwise, the particle can only tunnel downwards through the potential barrier.

A wavefunction $\tilde{\psi}(x,y,z)$ that obeys the free KGE in the conformal spacetime, and is hence supposed to have unlimited access to all the regions of such a space, appears to be restrained in the Minkowski spacetime. In the latter, either (i) the particle is confined to a restricted region of space where it has a quantized energy, or (ii) it propagates freely above the origin while the region below the origin is accessible to it only through tunneling, or (iii) it propagates freely throughout space provided that its energy along the $z$-direction exceeds a certain threshold. This is a very interesting physical result for the following three reasons.

First, recall that when dealing with the conformal transformation of the classical geodesic equation (evoked when arguing for the physical equivalence of the two spacetimes \cite{Dicke,Veiled}), one might argue that the physics appears different only because the particle's mass appears to be position-dependent. Within quantum mechanics, however, we see that one can hardly argue for the physical equivalence of the two spacetimes. Indeed, Dicke's rescaled-clocks-and-rulers argument \cite{Dicke} does not help us understand why a continuous energy spectrum becomes suddenly quantized or why a certain region of space suddenly becomes forbidden for a quantum particle (see, e.g., Refs.\,\cite{Debate1,Debate2,Debate3,Debate4,Debate5,Debate6,Debate7,Debate8,Debate9,Debate10,Debate11,Debate12,Debate13} for other contributions to such an ongoing debate). 

Second, it is not a trivial task to justify the emergence of a repulsive gravitational potential when we know that gravity is, in contrast to electromagnetism, a force that is always attractive. By deforming spacetime with our conformal factor $\Omega(z)$, a ``repulsive'' gravity, much like the one arising from the expansion of the Universe, emerges. In contrast to the cosmological case, however, we created a potential barrier for the particle with a deformed spacetime that reduces, far from the origin $z=0$, to the flat Minkowski spacetime. 

The last, but not the least, interesting feature of our result in this section comes to light when referring to the nearly century-old unsettled debate in the foundations of quantum mechanics: the $\psi$-epistemic vs. the $\psi$-ontic nature of the quantum state\footnote{See Ref.\,\cite{OnticReview} for a nice review on such a distinction and Refs.\,\cite{Ontic+Epistemic1,Ontic+Epistemic2} for a more recent discussion on the subtleties behind such a classification.}. In fact, if we accept, {\it \`a la} Dicke, the physical equivalence of the two spacetimes, the observer should conclude that the wavefunction cannot be epistemic in nature, {\it i.e.}, it cannot merely represent one's knowledge about a system. For it were so, the measured observables of the system would have appeared invariant just as the observer's clocks and rulers do. On the other hand, if we accept the physical non-equivalence of the two spacetimes, it is hard to understand the existence of the transformation (\ref{PhiNonlocality}) when assigning a $\psi$-epistemic character to the wavefunction. Only a $\psi$-ontic character of the wavefunction, {\it i.e.,} by assigning a physical meaning to the latter, would naturally explain the nonlocal physical influence of spacetime on the wavefunction.
\subsection{Linking the Schr\"odinger wavefunctions}\label{Subsec:Schrodinger}
To obtain a probability density in the nonrelativistic approximation from the time-components of the Klein-Gordon currents, we multiply both sides of Eq.\,(\ref{PhiNonlocality}) by $i\Omega^2\hbar/2mc$ as well as by $\sqrt{\gamma}$, where $\gamma$ is the determinant of the induced 3-metric $\gamma_{ab}$ on the spatial slice of spacetime. Using the ansatz $\varphi(x)=\psi(x)e^{-\frac{i}{\hbar}mc^2t}$ and the fact that $i\hbar\partial_t\psi=\mathcal{E}\psi$, we extract from the right-hand side of the resulting equation the following nonrelativistic approximation for the probability density on the chosen spatial slice:
\begin{align}\label{OriginalDensity}
\frac{i\hbar\sqrt{\gamma}}{2mc^2}\left(\varphi^*\frac{\partial\varphi}{\partial t}-\varphi\frac{\partial\varphi^*}{\partial t}\right)&=\frac{\hbar\sqrt{\gamma}}{2mc^2}\left(\frac{2mc^2}{\hbar}\psi^*\psi+i\psi^*\frac{\partial\psi}{\partial t}-i\psi\frac{\partial\psi^*}{\partial t}\right)\nonumber\\
&\approx\frac{\sqrt{\gamma}\left(mc^2+\mathcal{E}\right)}{mc^2}\,\psi^*\psi\nonumber\\
&\approx\sqrt{\gamma}\,\psi^*\psi.
\end{align}
We recognize in this expression the genuine probability density $\psi^*\psi$ one builds from the Schr\"odinger wavefunction. The multiplicative factor $\sqrt{\gamma}$ takes care of converting the coordinate measure ${\rm d}^3x$ into a physical volume over which the probability density should be integrated. On the other hand, using the ansatz $\tilde{\varphi}(x)=\tilde{\psi}(x)e^{-\frac{i}{\hbar}\tilde{m}c^2t}$, and the fact that one would write $i\hbar\partial_t\tilde{\psi}=\tilde{\mathcal{E}}\tilde{\psi}$ in the conformal spacetime, the left-hand side of Eq.\,(\ref{PhiNonlocality}) yields
\begin{align}\label{ConfDensity}
\frac{i\hbar\sqrt{\gamma}\,\Omega^2}{2mc^2}\left(\tilde{\varphi}^*\frac{\partial\tilde{\varphi}}{\partial t}-\tilde{\varphi}\frac{\partial\tilde{\varphi}^*}{\partial t}\right)&=\frac{\hbar\sqrt{\gamma}\,\Omega}{2\tilde{m}c^2}\left(\frac{2\tilde{m}c^2}{\hbar}\tilde{\psi}^*\tilde{\psi}+i\tilde{\psi}^*\frac{\partial\tilde{\psi}}{\partial t}-i\tilde{\psi}\frac{\partial\tilde{\psi}^*}{\partial t}\right)\nonumber\\
&\approx\frac{\sqrt{\gamma}\,\Omega\left(\tilde{m}c^2+\tilde{\mathcal{E}}\right)}{\tilde{m}c^2}\,\tilde{\psi}^*\tilde{\psi}\nonumber\\
&\approx\sqrt{\gamma}\,\Omega\,\tilde{\psi}^*\tilde{\psi}.
\end{align}
In the first step, we have used $\tilde{m}=m/\Omega$. In the last step, we have used the fact that $\tilde{m}c^2\gg\tilde{\mathcal{E}}$, which follows from the nonrelativistic approximation in the original spacetime and from the similar way mass and energy transform. Equating the last line of Eq.\,(\ref{OriginalDensity}) and the last line of Eq.\,(\ref{ConfDensity}), leads to the identity, $\psi^*\psi=\Omega\,\tilde{\psi}^*\tilde{\psi}$. 

This suggests that under a Weyl transformation of spacetime, Schr\"odinger's wavefunction $\psi(x)$ is related to its counterpart in the conformal spacetime in a linear and local way according to $\tilde{\psi}(x)=\Omega^{-\frac{1}{2}}\psi(x)$. However, as indicated above, this apparent locality and linearity came at the price of selecting a preferred spatial slicing in both spacetimes. In fact, the time $t$ selected in the original spacetime to isolate the mass $m$ of the particle has a priori no justified reason to be the same as the one selected in the new spacetime to isolate the mass $\tilde{m}$. To see that this act actually leads to a serious fundamental physical issue, note that identifying the last line of Eq.\,(\ref{OriginalDensity}) with the last line of Eq.\,(\ref{ConfDensity}) leads to
\begin{equation}\label{UnitariyViolation}
    \int_V \sqrt{\gamma}\,\psi^*\psi\,{\rm d}^3x=\int_V \Omega^{-\frac{1}{2}}\sqrt{\tilde{\gamma}}\,\tilde{\psi}^*\tilde{\psi}\,{\rm d}^3x.
\end{equation}
On the right-hand side we have used the fact that $\tilde{\gamma}=\Omega^3\gamma$. The result (\ref{UnitariyViolation}) clearly violates the unitarity of the transformation, for both wavefunctions $\psi(x)$ and $\tilde{\psi}(x)$ have to be normalised using their respective integration measures.

The deep physical reason is the way we relied on the mass of the particle to select a preferred time to extract the Schr\"odinger equation. The issue (\ref{UnitariyViolation}), which is the main result of this section, can therefore be seen as another manifestation of the famous problem of time in quantum gravity. The remarkable thing, however, is that here we encountered such a problem not by trying to describe spacetime quantum mechanically, but as a result of an apparently innocent attempt to describe the general effect of spacetime on the wavefunction. Note that, had we chosen the time $t$ in the new spacetime based on the mass $m$ of the particle in the old spacetime, we would have had the same issue. 

We believe that this issue can be traced back to the problem of time in quantum gravity \cite{Isham,PLB2008} because the time $t$ we chose is tied up to the concept of mass. As the mass is tied up to the conformal factor shaping the new spacetime, any preferred choice of time would automatically amount to an unjustified preferred spatial slicing.
\section{In the de Broglie-Bohm approach}\label{dBBApproach}
In the de Broglie-Bohm (dBB) causal interpretation of quantum mechanics \cite{Broglie,Bohm,Holland}, a particle of mass $m$ moves in a deterministic way by following a trajectory $x^\mu(\tau)$ extracted from the Schr\"odinger equation by using the ansatz $\psi(x)=R(x)e^{\frac{i}{\hbar}S(x)}$ for the wavefunction $\psi(x)$. When extending the dBB approach to the relativistic regime \cite{Broglie1960}, one would have simply to insert the ansatz $\varphi(x)=R(x)e^{\frac{i}{\hbar}S(x)}$ into the KGE. The resulting equation then splits into a purely real equation and a purely imaginary equation that read, respectively, as follows:
\begin{align}
    g^{\mu\nu}\nabla_\mu S\nabla_\nu S&=-\left(m^2c^2-\frac{\hbar^2\Box R}{R}\right),\label{RealPart}\\
    \nabla_\mu\left(R^2\nabla^\mu S\right)&=0.\label{ImaginaryPart}
\end{align}
Because of the form of the first equation, which is reminiscent of the relativistic and conformally invariant equation $g_{\mu\nu}p^\mu p^\nu=-m^2c^2$, one is tempted to postulate, as it is done in the nonrelativistic regime, that the 4-velocity of the particle should be written as,
\begin{equation}\label{4Velocity}
u^\mu=-\frac{\nabla^\mu S}{m}\Longrightarrow\frac{{\rm d}u^\mu}{{\rm d}\tau}+\Gamma_{\nu\rho}^\mu u^\nu u^\rho=c^2g^{\mu\nu}\nabla_\nu\left(\frac{\hbar^2}{2m^2c^2}\frac{\Box R}{R}\right).
\end{equation}
The second identity in this equation is obtained by substituting the first into Eq.\,(\ref{RealPart}) and then taking the total derivative of the resulting equation with respect to an affine parameter $\tau$. The parentheses on the right-hand side of Eq.\,(\ref{4Velocity}) would then be interpreted as an effective, or a ``quantum'', mass term for the particle.

Unfortunately, since the term $\Box R$ is not in general positive-definite, such an interpretation encounters the issue of tachyonic solutions. Among the proposals \cite{Shojai,Nikolic,ApJ} to remedy this issue, there has been the suggestion to postulate that the equation of motion for the particle in curved spacetime should rather be written as \cite{Shojai},
\begin{equation}\label{ModifiedGeodesic}
    \frac{{\rm d}u^\mu}{{\rm d}\tau}+\Gamma_{\nu\rho}^\mu u^\nu u^\rho=\left(c^2g^{\mu\nu}-u^\mu u^\nu\right)\nabla_\nu\left(\frac{\hbar^2}{2m^2c^2}\frac{\Box R}{R}\right).
\end{equation}
This equation looks again like a corrected geodesic equation, but with a more involved force term on the right-hand side. The content inside the last parentheses is then thought of as being $\ln(\mathcal{M}/m)$, where $\mathcal{M}$ is thus interpreted as the ``quantum mass" for the particle by defining\footnote{Note that we chose to define here $\mathcal{M}$ with an extra factor of $\frac{1}{2}$ in the exponential, in contrast to the definition given in Ref.\,\cite{Shojai} where such a factor appears instead as a multiplicative constant on the right-hand side of Eq.\,(\ref{ModifiedGeodesic}). This does not alter our subsequent conclusion. The reason behind our choice is that it allows us to neatly see the analogy in the formalism ---and the difference in the physics--- of the more familiar effect of conformal transformations on the geodesic equation.} \cite{Shojai},
\begin{equation}\label{QuantumMass}
    \mathcal{M}\equiv m\exp\left(\frac{\hbar^2}{2m^2c^2}\frac{\Box R}{R}\right).
\end{equation}
The reason behind such a definition is that the geodesic equation (\ref{ModifiedGeodesic}) then takes the form $\mathcal{M}\ddot{u}^\mu=-\mathcal{M}\Gamma_{\nu\rho}^\mu u^\nu u^\rho+\left(c^2g^{\mu\nu}-u^\mu u^\nu\right)\nabla_\nu\mathcal{M}$ \cite{Shojai}, which looks like Newton's second law with two forces on the right-hand side: a gravitational force and a quantum force. Although the latter force, and hence also the quantum mass, come from the quantum nature of the particle ---\! as $\mathcal{M}$ is related to the amplitude $R(x)$ of $\varphi(x)$\! ---, it has been argued in Ref.\,\cite{Shojai} that the mass $\mathcal{M}$ should rather be thought of as a geometric ({\it i.e.}, a spacetime) effect.
This interpretation is suggested by the fact that such a term could be eliminated ---\! analogously to what de Broglie himself did for Eq.\,(\ref{RealPart}) in Ref.\,\cite{Broglie1960} \!--- by a conformal transformation of the metric with a conformal factor given by, $\Omega=\exp\left(\frac{\hbar^2}{2m^2c^2}R^{-1}\Box R\right)$. In this section, we examine the conformal behavior of both postulates (\ref{4Velocity}) and (\ref{ModifiedGeodesic}) under a Weyl transformation.

Let us start with the postulate of the 4-velocity as given by the first identity in Eq.\,(\ref{4Velocity}). We set $\varphi(x)=R(x)e^{\frac{i}{\hbar}S(x)}$ for the wavefunction satisfying the KGE in the original spacetime and $\tilde{\varphi}(x)=\tilde{R}(x)e^{\frac{i}{\hbar}\tilde{S}(x)}$ for the wavefunction satisfying the KGE in the conformal spacetime. Plugging these ansatzes into Eq.\,(\ref{PhiNonlocality}), and then multiplying both sides of the equation by $m^{-1}$, the equation yields,
\begin{equation}\label{SNonlocality}
R^2\frac{\nabla_\mu S}{m}=\Omega\tilde{R}^2\frac{\nabla_\mu \tilde{S}}{\tilde{m}}.
\end{equation}
First of all, this result shows that the phase $S(x)$ of the wavefunction of the particle does not transform independently of its amplitude $R(x)$. Second, if one wishes to keep postulating that the 4-velocity of the particle should solely be expressed in terms of the phase $S(x)$ as given by the first identity in Eq.\,(\ref{4Velocity}), then a conformal transformation of geometry renders such a requirement ambiguous. Indeed, not only the conformal transformation $\tilde{u}^\mu=u^\mu/\Omega$ of a genuine 4-velocity is then not recovered\footnote{The transformation of $u^\mu$ comes from combining the definition $u^\mu={\rm d}x^\mu/{\rm d}\tau$ with the fact that the proper time transforms as ${\rm d}\tilde{\tau}=\Omega{\rm d}\tau$ while the coordinates $x^\mu$ are conformally invariant.}, but even the fact that the 4-velocity should a priori be independent of the amplitude $R(x)$ is lost. 

Yet, regardless of what one postulates for the 4-velocity of the particle, the interesting fact about the result (\ref{SNonlocality}) is that the phase of the particle transforms in a nonlocal way. This can easily be seen for the case of a constant amplitude $R(x)$ in the original spacetime ---\,like the case of a plane wave. For then, the phase $S(x)$ in the original spacetime is related to the phase $\tilde{S}(x)$ of the new spacetime as follows:
\begin{equation}\label{ConfS}
S(x)=\int\frac{\Omega^2\tilde{R}^2}{R^2}\partial_\mu \tilde{S}(x)\,{\rm d}x^\mu.
\end{equation}

Nonlocality is indeed much more evident from this result. Let us now examine the consequence of a conformal transformation of spacetime on the postulate (\ref{ModifiedGeodesic}). First, we have
\begin{equation}\label{ConfRealPart}
    \tilde{g}^{\mu\nu}\tilde{\nabla}_\mu \tilde{S}\tilde{\nabla}_\nu \tilde{S}=-\left(\tilde{m}^2c^2-\frac{\hbar^2\tilde{\Box}\tilde{R}}{\tilde{R}}\right).
\end{equation}
Eliminating the gradients of $S(x)$ and $\tilde{S}(x)$ from this equation and from Eq.\,(\ref{RealPart}) by using identity (\ref{SNonlocality}), yields
\begin{equation}\label{RNonlocal}
    \frac{\hbar^2}{2m^2c^2}\frac{\Box R}{R}-\frac{1}{2}=\frac{\Omega^4\tilde{R}^4}{R^4}\left(\frac{\hbar^2}{2\tilde{m}^2c^2}\frac{\tilde{\Box}\tilde{R}}{\tilde{R}}-\frac{1}{2}\right).
\end{equation}
Taking the exponential of both sides of this equation and then using the definition (\ref{QuantumMass}) of the quantum mass $\mathcal{M}$, we deduce that the latter is related to the quantum mass $\tilde{\mathcal{M}}$ of the conformal spacetime by
\begin{equation}\label{Mtransform}
\frac{\tilde{\mathcal{M}}}{\tilde{m}}=\left(\frac{\mathcal{M}}{m}\right)^{\frac{R^4}{\Omega^4\tilde{R}^4}}\exp\left(\frac{1}{2}-\frac{R^4}{2\Omega^4\tilde{R}^4}\right).
\end{equation}
We clearly see that not only the mass $\mathcal{M}$ does not transform the way a usual mass does under a conformal transformation, but it is, in addition, impossible to reduce such a transformation solely to a geometric effect. The latter is indeed supposed to manifest itself only in the form of a functional of the term $R^{-1}\Box R$, as was discussed below Eq.\,(\ref{QuantumMass}). This special mass depends not only on what was the amplitude $R$ of the wavefunction in the old spacetime, but it depends also on what would the new amplitude $\tilde{R}$ be in the new spacetime. This observation makes it hard to keep assigning to such a mass a purely geometric origin.

Finally, using Eqs.\,(\ref{RNonlocal}) and (\ref{QuantumMass}), as well as $\tilde{u}^\mu=u^\mu/\Omega$, ${\rm d}\tilde{\tau}=\Omega{\rm d}\tau$ and the way the Christoffel symbols $\Gamma_{\nu\rho}^\mu$ transform \cite{ConformalIssue}, we arrive at the following form for the geodesic equation in the conformal spacetime,
\begin{equation}\label{ConfModifiedGeodesic}
    \frac{{\rm d}\tilde{u}^\mu}{{\rm d}\tilde{\tau}}+\tilde{\Gamma}_{\nu\rho}^\mu\tilde{u}^\nu \tilde{u}^\rho=\left(c^2\tilde{g}^{\mu\nu}-\tilde{u}^\mu \tilde{u}^\nu\right)\left[\tilde{\nabla}_\nu\left(\frac{\Omega^4\tilde{R}^4}{R^4}\ln\frac{\tilde{\mathcal{M}}}{\tilde{m}}\right)-\tilde{\nabla}_\nu\left(\frac{\Omega^4\tilde{R}^4}{2R^4}\right)-\frac{\tilde{\nabla}_\nu\Omega}{\Omega}\right].
\end{equation}
The last term inside the square brackets in this expression is what emerges from the usual conformal transformation of the classical geodesic equation of a point particle. Unfortunately, neither the first term nor the second term inside the square brackets could be expressed solely in terms of the conformal quantum mass $\tilde{\mathcal{M}}$ and the conformal factor $\Omega$. The presence of the second term inside the square brackets makes things even worse. Such a term can, indeed, neither be interpreted as coming from the quantum mass nor as being due to pure geometry. This observation adds weight to the difficulties that already emerged from the result (\ref{Mtransform}) when attempting to interpret the quantum effects as being geometric in nature. 

The main conclusions of this section are: (i) the nonlocality of the transformation of the quantum phase is more apparent in the dBB approach, (ii) the Weyl transformation efficiently filters out non-geometric entities in the approach, and (iii) new light is shed on the issue of identifying the four-velocity with $-\nabla^\mu S/m$.
\section{Revisiting the conformal invariance of Maxwell's equations}\label{LinkWithMaxwell}
The conformal transformation of Maxwell's equations in curved spacetime is very well studied in the literature and it is well known that (in the 4-dimensional spacetimes we are interested in this paper) these equations are conformally invariant \cite{ConformalIssue,Particles+Fields,Wald,Fulton,Cunningham,Bateman,Rosen,Codirla,Cote}\footnote{The effect of the pure Weyl conformal transformation subtly emerges mixed with a diffeomorphism under a coordinate conformal transformation when the latter is expressed in curvilinear coordinates. This has been shown in detail for the case of Maxwell's equations in Ref.\,\cite{Rosen} by extending the results of Refs.\,\cite{Cunningham,Bateman} to curved spacetime.}. What has not been discussed in the literature, however, is where does the conformal invariance of Maxwell's equations stand with respect to the way massive matter waves transform under a Weyl rescaling of spacetime. Our goal in this section is thus to fill this gap in the literature. 

Let us then start by recalling Maxwell's equations in the presence of a charged current. The electromagnetic tensor is defined through the vector potential by $F_{\mu\nu}=\partial_\mu A_\nu-\partial_\nu A_\mu$. The first set of Maxwell's equations in the presence of a charged current density $j_\mu$ in curved spacetime is $\nabla^\mu F_{\mu\nu}=\mu_0j_\nu$, where $\mu_0$ is the permeability of free space. The second set of Maxwell's equations are just the Bianchi identities $\epsilon^{\sigma\mu\nu\rho}\nabla_{\mu}F_{\nu\rho}=0$, where $\epsilon^{\mu\nu\rho\sigma}$ is the totally antisymmetric tensor \cite{Wald}. For brevity, both sets of equations will be henceforth referred to as two single equations when there is no risk of confusion. After substituting the expression of $F_{\mu\nu}$ in terms of the derivatives of $A_\mu$ and rearranging the covariant derivatives thanks to the identity $\nabla^\mu\nabla_\nu A_\mu=\nabla_\nu\nabla^\mu A_\mu+{\mathcal{R}_\nu}^\mu A_\mu$ \cite{Wald}, the first Maxwell equation takes the form
\begin{equation}\label{Maxwell}
\Box A_\mu-\nabla_\mu\nabla^\nu A_\nu-{\mathcal{R}_\mu}^\nu A_\nu=\mu_0j_\mu.    
\end{equation}
Here, $\mu_0$ is the permeability of free space and $\mathcal{R}_{\mu\nu}$ is the Ricci tensor. The conformal invariance of the second Maxwell equation is guaranteed by the conformal invariance of the vector potential, $\tilde{A}_\mu=A_\mu$, that yields $\tilde{F}_{\mu\nu}=F_{\mu\nu}$. On the other hand, a straightforward calculation shows that the conformal invariance of the first Maxwell equation is also guaranteed since $\tilde{\nabla}^\nu\tilde{F}_{\mu\nu}=\Omega^{-2}\nabla^\nu F_{\mu\nu}$ \cite{Wald} and the charge current $j^\mu$ transforms as $\tilde{j}_\mu=\Omega^{-2}j_\mu$. To see that the latter is classically the case, recall that the 4-current density $j^\mu$ is defined by $\rho_e u^\mu$, where $\rho_e$ is the charge volume density and $u^\mu$ is the 4-velocity of the charge. Since $\tilde{u}^\mu=u^{\mu}/\Omega$, whereas the physical volume transforms like $\tilde{V}=\Omega^3 V$, one indeed arrives at $\tilde{j}_\mu=\Omega^{-2}j_\mu$ provided that electric charge $e$ is conformally invariant.

A couple of interesting remarks are in order. The first is that, in contrast to mass, to which a matter wave is associated, electric charge has no pure ``charge wave" associated to it. This can thus be viewed as the reason behind the conformal invariance of charge. The second remark is that Eq.\,(\ref{Maxwell}) remains conformally invariant if one includes the mass term $m^2A_\mu A^\mu$. That this is the case can immediately be verified by recalling that $\tilde{m}=m/\Omega$ and $\tilde{A}^\mu=\tilde{g}^{\mu\nu}\tilde{A}_\nu=A^\mu/\Omega^2$. This indicates that the conformal invariance of Maxwell's equations is not due to the fact that they describe massless particles, but due to the conformal invariance of the vector potential $A_\mu$ and to the fact that the latter couples to gravity in just the ``right" way to guarantee conformal invariance.

In the presence of an electromagnetic field minimally coupled to matter, the KGE in the original spacetime reads
\begin{equation}\label{KG+A}
\left[g^{\mu\nu}\left(\nabla_\mu+\frac{ie}{\hbar c}A_\mu\right)\left(\nabla_\nu+\frac{ie}{\hbar c}A_\nu\right)-\frac{m^2c^2}{\hbar^2}\right]\varphi=0.
\end{equation}
By going to the conformal spacetime, Eq.\,(\ref{KG+A}) takes the form
\begin{equation}\label{ConfKG+A}
\left[\tilde{g}^{\mu\nu}\left(\tilde{\nabla}_\mu+\frac{ie}{\hbar c}A_\mu\right)\left(\tilde{\nabla}_\nu+\frac{ie}{\hbar c}A_\nu\right)-\frac{2ie}{\hbar c}\frac{\tilde{\nabla}^\mu\Omega}{\Omega}A_\mu-\frac{\tilde{m}^2c^2}{\hbar^2}\right]\varphi=2\frac{\tilde{\nabla}^\mu\Omega}{\Omega}\tilde{\nabla}_\mu\varphi.
\end{equation}
We easily see from this equation that gauge invariance is preserved in the conformal spacetime as well. The scalar $\varphi(x)$ in the conformal spacetime would still have to transform under that gauge transformation as $\varphi\rightarrow e^{-\frac{ie}{\hbar c}\chi(x)}\varphi$. Now, in analogy with what we did to get Eq.\,(\ref{ConfKGCurrent}), we multiply both sides of Eq.\,(\ref{ConfKG+A}) by $\varphi^*$ and then we take the complex conjugate of the resulting equation and, finally, we subtract such an equation from the equation before complex conjugation, to arrive at the following result:
\begin{equation}\label{ConfKG+ACurrent}
\tilde{\nabla}^\mu\left[\Omega^{-2}\left(\varphi^*\tilde{\nabla}_{\mu}\varphi-\varphi\tilde{\nabla}_{\mu}\varphi^*+\frac{2ie}{\hbar c}A_\mu\varphi^*\varphi\right)\right]=0.
\end{equation}
Using the fact that the electric charge $e$ and the vector $A_\mu$ are both conformally invariant, together with the fact that a similar identity as Eq.\,(\ref{ConfKG+ACurrent}) ---\,but without the factor $\Omega^{-2}$ inside the square brackets\,--- is satisfied by $\tilde{\varphi}$ in the conformal spacetime, we conclude that,
\begin{equation}\label{PhiNonlocality+ACurrent}
\tilde{\varphi}^*\tilde{\nabla}_{\mu}\tilde{\varphi}-\tilde{\varphi}\tilde{\nabla}_{\mu}\tilde{\varphi}^*+\frac{2ie}{\hbar c}A_\mu\tilde{\varphi}^*\tilde{\varphi}=\Omega^{-2}\left(\varphi^*\tilde{\nabla}_{\mu}\varphi-\varphi\tilde{\nabla}_{\mu}\varphi^*+\frac{2ie}{\hbar c}A_\mu\varphi^*\varphi\right).
\end{equation}
By multiplying both sides of this identity by $i\Omega^2\hbar e/2m$, the right-hand side gives rise to the charged current density $j_\mu(x)$ with the right units for a current of charge:
\begin{equation}\label{Chargedj}
    j_\mu(x)=\frac{i\hbar e}{2m}\left(\varphi^*\nabla_{\mu}\varphi-\varphi\nabla_{\mu}\varphi^*+\frac{2ie}{\hbar c}A_\mu\varphi^*\varphi\right).
\end{equation}
On the other hand, using the fact that $\tilde{m}=m/\Omega$ and that electric charge is conformally invariant, when multiplied by $i\Omega^2\hbar e/2m$ the left-hand side of identity (\ref{PhiNonlocality+ACurrent}) gives rise to the product $\Omega\tilde{j}_\mu(x)$. In fact, the charged current density $\tilde{j}_\mu$ that naturally emerges from the solution $\tilde{\varphi}(x)$ to the KGE in the conformal spacetime reads
\begin{equation}\label{TildejToj}
    \tilde{j}_\mu(x)=\frac{i\hbar e}{2\tilde{m}}\left(\tilde{\varphi}^*\tilde{\nabla}_{\mu}\tilde{\varphi}-\tilde{\varphi}\tilde{\nabla}_{\mu}\tilde{\varphi}^*+\frac{2ie}{\hbar c}A_\mu\tilde{\varphi}^*\tilde{\varphi}\right)=\Omega^{-1}j_\mu(x).
\end{equation}
As a consequence, we just arrived at the conformal relation between the two charged current densities, $\tilde{j}_\mu(x)=\Omega^{-1}j_\mu(x)$. This is clearly in conflict with the conformal invariance of the first Maxwell equation in the presence of matter as we just saw. This is not what is expected from our classical argument based on the deterministic classical definition of the current $j^\mu=\rho_e u^\mu$ either. 

The reason behind such an issue becomes actually much clearer when using the Lagrangian approach. The Lagrangian density from which both equations emerge in Minkowski spacetime \cite{Greiner} gives rise to the following action in curved spacetime:
\begin{equation}\label{Action}
\mathcal{S}=\int\sqrt{-g}\left[-\frac{1}{4\mu_0}F_{\mu\nu}F^{\mu\nu}-\frac{\hbar^2}{2m}\left(\mathcal{D}_\mu^*\varphi^*\mathcal{D}^\mu\varphi+\frac{m^2c^2}{\hbar^2}\varphi^*\varphi\right)\right]{\rm d}^4x   
\end{equation}
Here, we defined as usual the minimal-coupling operator in curved spacetime to be $\mathcal{D}_\mu=\nabla_\mu-\frac{ie}{\hbar c}A_\mu$. The variation of this action with respect to $\varphi^*(x)$ yields the minimally-coupled KGE (\ref{KG+A}). The variation of the action with respect to the vector potential $A_\mu$ gives rise to the first Maxwell equation in the original spacetime:
\begin{equation}\label{MaxwellKG}
    \nabla^\nu F_{\mu\nu}=\mu_0\frac{i\hbar e}{2m}\left(\varphi^*\nabla_{\mu}\varphi-\varphi\nabla_{\mu}\varphi^*+\frac{2ie}{\hbar c}A_\mu\varphi^*\varphi\right).
\end{equation}
We indeed recognize on the right-hand side of this equation the charged current (\ref{Chargedj}). The question that we ask now is what should be the analog of the action integral (\ref{Action}) in the conformal spacetime? At first thought, the answer would be to simply rewrite the action with all its terms decorated with tildes, leaving only $e$ and the fundamental constants unaffected. That this option is not satisfactory can be seen by taking the variation of such an action with respect to $\tilde{A}_\mu$. Indeed, the result would then simply be $\tilde{\nabla}^\nu\tilde{F}_{\mu\nu}=\mu_0\tilde{j}_\mu$. This identity is not satisfactory, for when substituting into it $\tilde{\nabla}^\nu\tilde{F}_{\mu\nu}=\Omega^{-2}\nabla^\nu F_{\mu\nu}$ and then comparing it to Eq.\,(\ref{MaxwellKG}) we arrive at $\tilde{j}_\mu=\Omega^{-2}j_\mu$, which is in conflict with Eq.\,(\ref{TildejToj}). That this guess for the action integral in the conformal spacetime leads to the same issue encountered when considering the classical current $j^\mu=\rho_eu^\mu$ can be understood as follows.

Thanks to the definition (\ref{Chargedj}) of the current, we see that the middle term inside the square brackets in the action (\ref{Action}) can actually be written as $-j_\mu A^\mu$. This term represents the coupling between the charged current and electromagnetism just as one would have introduced such an interaction term in the classical case where $j^\mu=\rho_e u^\mu$. Therefore, adopting exactly the same action (\ref{Action}) in the conformal spacetime by simply decorating all its terms with tildes amounts to introducing in the conformal spacetime an interaction term of the form $-\tilde{j}_\mu\tilde{A}^\mu$. That this would not be the correct term for an interaction in the conformal spacetime can be seen from the fact that the current $\tilde{j}_\mu$ as given by Eq.\,(\ref{TildejToj}) is not conserved in the conformal spacetime. What is conserved in the conformal spacetime is the product $\Omega^{-1}\tilde{j}_\mu$. As a consequence, the equation $\tilde{\nabla}^\nu\tilde{F}_{\mu\nu}=\mu_0\tilde{j}_\mu$ one arrives at with the coupling term $-\tilde{j}_\mu\tilde{A}^\mu$ is inconsistent, for we easily check that $\tilde{\nabla}^\mu\tilde{\nabla}^\nu\tilde{F}_{\mu\nu}=0\neq \mu_0\tilde{\nabla}^\mu \tilde{j}_\mu$. 

From this observation, we conclude that the interaction term that one needs to use in an action integral in the conformal spacetime is a term of the form $-\Omega^{-1}\tilde{j}_\mu\tilde{A}^\mu$. The action integral in the conformal spacetime would then read as follows:
\begin{equation}\label{ConfAction}
\tilde{\mathcal{S}}=\int\sqrt{-\tilde{g}}\left[-\frac{1}{4\mu_0}\tilde{F}_{\mu\nu}\tilde{F}^{\mu\nu}-\frac{\hbar^2}{2\tilde{m}\Omega}\left(\tilde{\mathcal{D}}_\mu^*\tilde{\varphi}^*\tilde{\mathcal{D}}^\mu\tilde{\varphi}+\frac{\tilde{m}^2c^2}{\hbar^2}\tilde{\varphi}^*\tilde{\varphi}\right)\right]{\rm d}^4x.   
\end{equation}
This displays exactly the same structure of the Lagrangian for $\varphi$ we arrived at in Sec.\,\ref{UsingLagragian} in the absence of electromagnetism. This is therefore a satisfactory result as it shows the consistency of our analysis. 

Remarkably, this whole issue is very reminiscent of the issue one encounters when conformally transforming Einstein's field equations. In fact, the latter, which take the form $G_{\mu\nu}=T_{\mu\nu}$ in the natural units $8\pi G=c=1$ (where $G_{\mu\nu}$ is the Einstein purely geometric tensor) do not get transformed into $\tilde{G}_{\mu\nu}=\tilde{T}_{\mu\nu}$. Instead, the equations in the conformal spacetime take the form $\tilde{G}_{\mu\nu}=\Omega^2\tilde{T}_{\mu\nu}+T^\Omega_{\mu\nu}$ whenever the energy-momentum tensor of matter transforms like $\tilde{T}_{\mu\nu}=\Omega^{-2}T_{\mu\nu}$. The tensor $T^\Omega_{\mu\nu}$ is interpreted as an induced energy-momentum tensor \cite{ConformalIssue}. What we just witnessed here is a very similar pattern. The first Maxwell's equations do not transform into $\tilde{\nabla}^\nu\tilde{F}_{\mu\nu}=\mu_0\tilde{j}_\mu$, but rather into $\tilde{\nabla}^\nu\tilde{F}_{\mu\nu}=\mu_0\Omega^{-1}\tilde{j}_\mu$, whenever the charged current density $j_\mu$ is extracted from the KGE. The fact that the scalar $\varphi(x)$ does not couple the same way to spacetime geometry as the vector $A_\mu$ does is actually what makes Maxwell's equations in the presence of matter behave under a Weyl transformation as Einstein's equations do.
\section{A possible vector construction for the non-minimal coupling of $\varphi(x)$ with gravity}\label{Sec:VectorConstruction}

While the KGE is not invariant under a conformal transformation, it is possible to render the equation conformally invariant by adding one extra term, proportional to the Ricci scalar $\mathcal{R}$, that makes $\varphi(x)$ couple non-minimally to gravity:
\begin{equation}\label{KG6}
\left(g^{\mu\nu}\nabla_\mu\nabla_\nu-\frac{m^2c^2}{\hbar^
2}-\frac{\mathcal{R}}{6}\right)\varphi=0.
\end{equation}
This equation is conformally invariant when the transformation of the mass $m$ in the equation is properly taken into account \cite{ConformalIssue}. That extra term also arises in quantum field theory from a counter-term in the Lagrangian that renormalizes a specific theory with an interacting scalar field in curved spacetime \cite{Bunch}.

One cannot help but wonder whether this extra term could somehow find a simple origin in the conformal invariance of the sourceless Maxwell vector equation. What makes this quest legitimate is the presence of the Ricci tensor in Eq.\,(\ref{Maxwell}). In addition, just like Eq.\,(\ref{KG6}), even after associating a mass to $A_\mu$ which thus becomes the Proca massive vector field \cite{Proca}, Eq.\,(\ref{Maxwell}) remains conformally invariant.

Inspired by the conformal invariance of Eq.\, (\ref{Maxwell}), the possibility that comes to mind is to build a vector potential $A_{\mu}$ from the scalar $\varphi(x)$ by introducing a vector field $\xi_\mu$ such that $A_\mu=\varphi\,\xi_\mu$. On the other hand, we know that for Eq.\,(\ref{KG6}) to be conformally invariant $\varphi$ needs to transform as $\tilde{\varphi}(x)=\Omega^{-1}\varphi(x)$, whereas $A_\mu$ is conformally invariant. Therefore, the vector field $\xi_\mu$ needs to transform as $\tilde{\xi}_\mu=\Omega\xi_\mu$. Since the dynamics of the scalar $\varphi(x)$ should not be tied up to the dynamics of any extra field apart from the one arising from spacetime itself, the natural vector species that comes to mind is the set of tetrad vectors of spacetime $e_\mu^a$. The tetrad field does indeed transform as required \cite{ParallelBH}, $\tilde{e}^a_\mu=\Omega e^a_\mu$, because it is related to the metric of spacetime by $g_{\mu\nu}=\eta_{ab}e^a_\mu e^b_\nu$. To recover a curved-spacetime vector $\xi_\mu$ from the tetrad, we would then only need to project the tetrad along a given tangent-space direction $\xi^a$ according to $\xi_\mu=\xi_a e^a_\mu$. 

Unfortunately, setting $A_\mu=\varphi\,\xi_ae^a_\mu$ requires extra unwanted features, such as an extra constraint on the tangent-vector $\xi^\mu$ which, in turn, leads to a constraint on the spacetime tetrad itself. The latter requirement should, to all costs, be avoided as our goal is to build a Lagrangian for $\varphi$ independently of the dynamics of the spacetime hosting such a scalar field. Therefore, instead of introducing the extra Lorentz vector $\xi_a$, we are going to exploit the fact that the action contains only terms involving two $A$'s. As such, we never need a vector $\xi^a$ to be contracted with the Lorentz index of the tetrad in the first place. In fact, we would only need then to contract the tetrads via their Lorentz indices.

It is known that more terms can be added to the Maxwell action and, hence, also to the Proca action \cite{JCAP2010}. We shall choose a Lagrangian which leads to, at most second-order, field equations for the vector potential $A_\mu$. Therefore, our covariant Lagrangian takes the following form:
\begin{equation}\label{GeneralizedMxawell}
\mathcal{L}\!=\!c_1\nabla_\mu A_\nu\nabla^\mu A^\nu\!+\!c_2\nabla_\mu A_\nu\nabla^\nu A^\mu\!+\!c_3\nabla_\mu A^\mu\nabla_\nu A^\nu\!+\!c_4\varepsilon^{\mu\nu\rho\sigma}\nabla_\mu A_\nu\nabla_\rho A_\sigma\!+\!c_5A_\mu A^\mu
\!+\!c_6\nabla_\mu A\nabla^\mu A.
\end{equation}
Here, $A^2=A_\mu A^\mu$ and the constants $c_1,..., c_4$ together with $c_6$ are all arbitrary and dimensionless, whereas the constant $c_5$ has the dimensions of an inverse distance squared. $c_5$ should then be proportional to $m^2c^2/\hbar^2$. The symbol $\varepsilon^{\mu\nu\rho\sigma}$ stands for the totally antisymmetric Levi-Civita symbol. Note that we did not include the term $\nabla_\mu A^\mu$, for in addition of being a total derivative, such a term would not make it possible to contract the Lorentz index of the single tetrad.

Introducing now inside the Lagrangian (\ref{GeneralizedMxawell}) contracted tetrads weighted by the scalar field $\varphi$, the Lagrangian takes the following form after rearranging its terms by discarding total derivatives as they do not contribute to the equations of motion:
\begin{align}\label{GeneralizedPhiLagrangian}
\mathcal{L}&=c_1\nabla_\mu \left(\varphi\, e^a_\nu\right)\nabla^\mu\left(\varphi\, e_a^\nu\right)+c_2\nabla_\mu \left(\varphi\, e^a_\nu\right)\nabla^\nu\left(\varphi\, e_a^\mu\right)+c_3\nabla^\mu \left(\varphi\, e^a_\mu\right)\nabla_\nu\left(\varphi\, e_a^\nu\right)\nonumber\\
&\quad+4c_5\varphi^2+c_4\varepsilon^{\mu\nu\rho\sigma}\nabla_\mu\left(\varphi\,e_{a\nu}\right)\nabla_\rho\left(\varphi\,e^a_{\sigma}\right)+c_6\nabla_\mu\varphi\nabla^\mu\varphi\nonumber\\
&=\left(4c_1+c_2+c_3+c_6\right)\nabla_\mu\varphi\nabla^\mu\varphi+4c_5\varphi^2\nonumber\\
&\quad-\left[c_1e^\mu_a\Box e^a_\mu+\left(c_2+c_3\right)e^\mu_a\nabla_\mu\nabla^\nu e^a_\nu+c_2\mathcal{R}+c_4\varepsilon^{\mu\nu\rho\sigma}e^a_\sigma\nabla_\rho\nabla_\mu e_{a\nu}\right]\varphi^2.
\end{align}
The Ricci scalar $\mathcal{R}$ appeared in this expression thanks to the identity $\nabla_\mu\nabla_\nu\xi^\mu=\nabla_\nu\nabla_\mu\xi^\mu+{\mathcal{R}_\mu}^\nu\xi_\nu$ satisfied by any spacetime-valued vector $\xi^\mu$ \cite{Wald}. We have also used the property $e_{a\mu}e^a_\nu=g_{\mu\nu}$, which, in turn, implies that $e_{a\mu}e^{a\mu}=4$, where $e_a^\mu$ is the inverse tetrad such that $e^a_\mu e_a^\nu=\delta_\mu^\nu$ \cite{Tetrads}. We already see from this expression the appearance of the main terms of the non-minimally coupled scalar field's Lagrangian that leads to Eq.\,(\ref{KG6}) by variation with respect to $\varphi$. However, in order to obtain the sought-after scalar field's Lagrangian that leads exactly to Eq.\,(\ref{KG6}), we only need to set $c_1=c_4=0$, $c_2=-c_3$, $4c_5=-\frac{3}{2}m^2c^2/\hbar^2$, $c_2=\frac{1}{4}$ and $c_6=-\frac{3}{2}$. The Lagrangian reduces then to the following expression:
\begin{equation}\label{FinalLagrangian}
\mathcal{L}=-\frac{3}{2}\left(\nabla_\mu\varphi\nabla^\mu\varphi+\frac{m^2c^2}{\hbar^2}\varphi^2+\frac{R}{6}\varphi^2\right).
\end{equation}
The variation of this Lagrangian with respect to $\varphi$ gives exactly the KGE (\ref{KG6}). This confirms that the Weyl invariance of Maxwell's equations has to do with the special way the vector potential couples to spacetime geometry. This is the main result of this section as it shows that attaching a scalar field to the spacetime tetrads gives rise to a conformally invariant Lagrangian for that scalar field as well.                    

Note that a few more possible options for achieving such a final Lagrangian by starting from Eq.\,(\ref{GeneralizedPhiLagrangian}) exist. The first option would be to set $e^a_\mu=\partial_\mu\chi^a$ for some Lorentz-valued vector field $\chi^a$. Such a tetrad allows indeed to eliminate the unwanted terms inside the square brackets in Eq.\,(\ref{GeneralizedPhiLagrangian}). Unfortunately, such a choice leads to a flat spacetime when evaluating the Riemann tensor in terms of tetrads \cite{Tetrads}. The second possibility is to impose on the tetrads a constraint of the form $\nabla_\mu e_\nu^a+\nabla_\nu e_\mu^a=\frac{1}{2}g_{\mu\nu}\nabla^\rho e^a_\rho$. Unfortunately, such a constraint cannot represent a mere gauge fixing of the diffeomorphism freedom because it involves a total of $10$ constraints on the tetrads.

The third option is to use a constant-norm vector field $\xi_\mu$ without even invoking the tetrads of spacetime, so that $A_\mu=\varphi\xi_\mu$. However, this would work provided that either the vector field $\xi^\mu$ satisfies the first Maxwell's equation in vacuum, or that it is a conformal Killing vector field that satisfies $\nabla_\mu\xi_\nu+\nabla_\nu\xi_\mu=\frac{1}{2}g_{\mu\nu}\nabla_\rho\xi^\rho$. Unfortunately, both choices also require to have $\xi^\mu\xi^\mu R_{\mu\nu}=\alpha R$, for some constant $\alpha$. As the latter condition cannot be satisfied for arbitrary spacetimes, the validity of the non-minimally coupled KGE that would be derived from such a vector construction would be restricted to specific spacetimes only.

\section{Consequences on the Aharonov-Bohm effect}\label{AharonovBohm}
It is well known that the phase difference in the Aharonov-Bohm effect \cite{AB1,AB2} can be expressed in terms of the flux of $F_{\mu\nu}$ through a 2-surface area. Given the conformal  invariance of $F_{\mu\nu}$ and the conformal {\it noninvariance} of a surface area, one expects that a Weyl transformation would certainly alter the Aharonov-Bohm effect in the conformal spacetime. Here, we show that this is surprisingly not the case.

The Aharonov-Bohm effect is an interference effect that arises due to the two different paths a charged particle could take, around a solenoid for example, when propagating from the source to the detector \cite{AB2}. To find the phase difference $\delta\Phi_{AB}$ responsible for the effect, one has to subtract the two accumulated phases along each of the two paths. This amounts thus to finding the circulation of $A_\mu$ around a closed path. Using Stokes' theorem, the phase difference $\delta\Phi_{AB}$ through a closed path in the Aharonov-Bohm effect is given by\footnote{See Refs.\,\cite{Singleton1,Singleton2} for another interesting physical discussion on these covariant integrals.}
\begin{equation}\label{PhiAB}
    \delta\Phi_{AB}=\frac{e}{\hbar}\oint_CA_\mu{\rm d}x^\mu=-\frac{e}{2\hbar}\int_SF_{\mu\nu}{\rm d}x^\mu\wedge{\rm d}x^\nu.
\end{equation}
Here, the closed line integral is performed over the closed path $C$ whereas the second integral is performed over the surface area $S$ bounded by the closed path, and $\wedge$ is the usual wedge product. In order to express the integrals in Eq.\,(\ref{PhiAB}) in terms of the line element ${\rm d}s$ along the closed path $C$ and the 2-surface element ${\rm d}S$ along the surface $S$, respectively, we need to introduce a convenient tetrad as follows \cite{Synge1,Synge2}. Choose a unit vector $t_{(1)}^\mu$ to be pointing along the closed path in the positive sense assigned to the loop. Choose a unit vector $t_{(2)}^\mu$ to be orthogonal to $t_{(1)}^\mu$, to be pointing along the outward normal to the closed path $C$ and to lie together with $t_{(1)}^\mu$ along the surface $S$. With such vectors, the phase difference (\ref{PhiAB}) takes the following form \cite{Synge1}:
\begin{equation}\label{CurvedPhiAB}
    \delta\Phi_{AB}=\frac{e}{\hbar}\oint_CA_\mu t_{(1)}^\mu{\rm d}s=\frac{e}{\hbar}\int_SF_{\mu\nu}t_{(1)}^\nu t_{(2)}^\mu{\rm d}S.
\end{equation}
Using now the fact that both $t_{(1)}^\mu$ and $t_{(2)}^\mu$ are unit spacelike vectors, we have $g_{\mu\nu}t_{(1)}^\mu t_{(1)}^\nu=1$ and $g_{\mu\nu}t_{(2)}^\mu t_{(2)}^\nu=1$. Since conformal transformations preserve the light-cone structure of spacetime, we deduce that our unit vectors transform as $\tilde{t}_{(1)}^\mu=\Omega^{-1}t_{(1)}^\mu$ and $\tilde{t}_{(2)}^\mu=\Omega^{-1}t_{(2)}^\mu$. On the other hand, the line element transforms as ${\rm d}\tilde{s}=\Omega{\rm d}s$ while the 2-surface element transforms as ${\rm d}\tilde{S}=\Omega^2{\rm d}S$. Combining these, together with the conformal invariance of both $A_\mu$ and $F_{\mu\nu}$, we immediately check the conformal invariance of the phase difference (\ref{CurvedPhiAB})\footnote{Note the similarity of Eq.\,(\ref{CurvedPhiAB}) with the flux integral $\delta M=\int{\rm d}\lambda\int_ST_{\mu\nu}\xi^\mu\xi^\nu{\rm d}S$ from the first law of black hole mechanics. While the former is conformally invariant, the latter is not \cite{BHThermo}.}.

This is the main result of this section. We have showed that although the Aharonov-Bohm effect (i) involves an integrated flux over a conformally noninvariant surface area and (ii) it arises from a phase difference while Eq.\,(\ref{ConfS}) implies a nonlocal link between phases in two conformal spacetimes, the effect itself is conformally invariant. This invariance can thus be understood as due to the topological nature of the effect and that conformal transformations preserve the topology of spacetime. 
\section{Summary and discussion}
We have examined the conformal noninvariance of the KGE under its many different aspects. We started with the fully relativistic regime and showed that both the Lagrangian formalism and the analysis based on the current conservation lead to the same interpretation. The noninvariance of the equation and the nonlocality of the transformation of the wavefunction are due to the interaction of the latter with a geometrically-induced current. 
In addition, our results are not limited to the linear KGE, for they hold even when considering the nonlinear Gross-Pitaevskii equation in curved spacetime. In all cases, the conformal transformation of mass was found to be central. Such a transformation is not borrowed from classical physics alone, but is itself intimately related to quantum mechanics coupled to gravity since it is derivable from the Weyl transformation of the Compton wavelength.

We saw that the KGE coupled to electromagnetism does not allow Maxwell's equation to transform the way it does with a classical current. The fact that we treated Maxwell's equation classically cannot be the origin of the issue, for we treated $\varphi(x)$ as a mere classical complex scalar field too. Its quantum nature was invoked only to make contact with the concept of probability density in quantum mechanics. As such, no semi-classical approach (in which the current $j^\mu$ would be replaced by its expectation value $\braket{j^\mu}$) or a fully quantum analysis (in which both $A_\mu$ and $\varphi$ would be subject to second quantization) were required. A quantum field theoretical analysis would be the next step for extending our present results.        

Having restricted our analysis to the first quantized Klein-Gordon field has also allowed us to make contact with the Schr\"odinger equation and its wavefunction. The issue we discovered is the violation of the unitarity of quantum mechanics by a mere passage from one spacetime to another. We argued that the issue could be related to the problem of time in quantum gravity. We also learned that Weyl transformations pose serious challenges to the epistemic interpretation of the wavefunction. 

We then examined the de Broglie-Bohm approach to quantum mechanics in light of the conformal noninvariance of the KGE. We came to the conclusion that a possible link between geometry and the quantum, as suggested in recent works in the literature, is still far from being convincing within such an approach.

Next, inspired by the conformal invariance of Maxwell's equations, we sought a possible vector construction for the conformally invariant KGE. We saw that such a construction is indeed possible by attaching a scalar field to the tetrads of spacetime. We then discussed the conformal invariance of the Aharonov-Bohm effect that arises from the gauge invariance of the KGE. We explained the unexpected invariance of the flux that leads to the effect despite the conformal noninvariance of the surface area that defines the flux, as being due to the topological nature of the effect and that spacetime topology is preserved under conformal transformations. 

It is remarkable that so much interesting physics that begs for further investigation arises from an otherwise mundane conformal noninvariance of an equation.

{\section*{Acknowledgments}
The authors are grateful to the anonymous referee for the constructive criticism and comments that helped improve the clarity of our manuscript. This work is supported in part by the SRC Interdisciplinary Team Grant from Bishop's University and by the Natural Sciences and Engineering Research Council of Canada (NSERC) Discovery Grant No. RGPIN-2017-05388.}

\begin{appendices}
\section{Derivation of the KGE from the Lagrangian (\ref{TildePhiLagrangian})}\label{Appendix}
By varying the Lagrangian (\ref{TildePhiLagrangian}) with respect to $\tilde{\varphi}^*$, we find
\begin{equation}
    \delta\tilde{\mathcal{L}}=-\frac{\hbar^2}{2m}\tilde{\nabla}_\mu\left(\delta \tilde{\varphi}^*\tilde{\nabla}^\mu\tilde{\varphi}\right)+\frac{\hbar^2}{2m}\left(\tilde{\nabla}_\mu\tilde{\nabla}^\mu\tilde{\varphi}-\frac{\tilde{m}^2c^2}{\hbar^2}\tilde{\varphi}\right)\delta\tilde{\varphi}^*.
\end{equation}
We have replaced $\tilde{m}\Omega$ by the constant $m$ in the denominator. Requiring that $\delta\tilde{\mathcal{L}}$ vanishes for an arbitrary variation $\delta\tilde{\varphi}^*$ in the bulk, the content of the second pair of parentheses has to vanish, yielding the KGE in the conformal spacetime. Multiplying the KGE on the left by $\varphi^*$ then leads to,
\begin{align}
    &\tilde{\varphi}^*\tilde{\nabla}_\mu\tilde{\nabla}^\mu\tilde{\varphi}-\frac{\tilde{m}^2c^2}{\hbar^2}\tilde{\varphi}^*\tilde{\varphi}=0\nonumber\\
    \Rightarrow&\tilde{\nabla}_\mu\left(\tilde{\varphi}^*\tilde{\nabla}^\mu\tilde{\varphi}-\tilde{\varphi}\tilde{\nabla}^\mu\tilde{\varphi}^*\right)=0\nonumber\\
    \Rightarrow&\tilde{\nabla}_\mu\left[\Omega^{-2}\left(\varphi^*\tilde{\nabla}^\mu\varphi-\varphi\tilde{\nabla}^\mu\varphi^*\right)\right]=0\nonumber\\
    \Rightarrow&-2\frac{\tilde{\nabla}^\mu\Omega}{\Omega}\left[\Omega^{-2}\left(\varphi^*\tilde{\nabla}_\mu\varphi-\varphi\tilde{\nabla}_\mu\varphi^*\right)\right]+\Omega^{-2}\left(\varphi^*\tilde{\nabla}_\mu\tilde{\nabla}^\mu\varphi-\varphi\tilde{\nabla}_\mu\tilde{\nabla}^\mu\varphi^*\right)=0\nonumber\\
    \Rightarrow&-2\frac{\tilde{\nabla}^\mu\Omega}{\Omega}\left(\Omega^{-2}\varphi^*\tilde{\nabla}_\mu\varphi\right)+\Omega^{-2}\left(\varphi^*\tilde{\nabla}_\mu\tilde{\nabla}^\mu\varphi-\Omega^{-2}\varphi\nabla_\mu\nabla^\mu\varphi^*\right)=0\nonumber\\
    \Rightarrow&-2\frac{\tilde{\nabla}^\mu\Omega}{\Omega}\left(\Omega^{-2}\varphi^*\tilde{\nabla}_\mu\varphi\right)+\Omega^{-2}\left(\varphi^*\tilde{\nabla}_\mu\tilde{\nabla}^\mu\varphi-\frac{m^2c^2}{\Omega^2\hbar^2}\varphi\,\varphi^*\right)=0\nonumber\\
    \Rightarrow&\left(\tilde{\nabla}_\mu\tilde{\nabla}^\mu-\frac{\tilde{m}^2c^2}{\hbar^2}\right)\varphi=2\frac{\tilde{\nabla}^\mu\Omega}{\Omega}\tilde{\nabla}_\mu\varphi.
\end{align}
In the second step, we subtracted the first line from its complex conjugate. In the third step, we used identity (\ref{PhiNonlocality}). In the fifth step, we used the geometric identity $\tilde\nabla^\mu\tilde\nabla_\mu-2\Omega^{-1}\tilde{\nabla}^\mu\Omega\tilde{\nabla}_\mu=\Omega^{-2}\nabla^\mu\nabla_\mu$. In the sixth step, we used the KGE extracted from the Lagrangian (\ref{PhiLagrangian}). In the last step, we used $m/\Omega=\tilde{m}$.

\end{appendices}

\end{document}